# Nanostructured Zinc Oxide as a Prospective Room Temperature Thermoelectric Material

Pawan Kumar[1], M. Kar[1], Anup V. Sanchela[2], C. V. Tomy[2] and Ajay D. Thakur[1]

[1] *Department of Physics, Indian Institute of Technology Patna, Patna-800013, India.*
[2] *Department of Physics, Indian Institute of Technology Bombay, Mumbai-400076, India.*

**Abstract.** Nanostructured Zinc oxide (ZnO) was synthesized via a ball milling for 10 hours using high energy planetary ball mill. Phase purity and homogeneity of all the samples have been investigated by X-ray diffraction (XRD) and Field Emission Scanning Electron Microscopy (FE-SEM). All the diffraction peaks can be indexed to the hexagonal phase ZnO with hexagonal symmetry (space group $P6_3mc$). Average crystallite size was observed to be 20 nm. There was a remarkable suppression in thermal conductivity ( $\kappa$ ) compared to the bulk values by a factor of ~50 at room temperature. This suggests to the possibility of using nanostructured ZnO as a prospective room temperature thermoelectric material.

## INTRODUCTION

ZnO is one of the most important thermoelectric materials and has extensive applications due to its unique properties such as high carrier mobility and Seebeck coefficient [1]. There were very few facile and effective top-down methods developed to adjust or optimize the carrier concentration of thermoelectric nanomaterials. Thermoelectric performance of the materials, which is characterized by a dimensionless figure of merit $zT = \sigma S^2 T/\kappa$, where $\sigma$ is the electrical conductivity, $S$ is the thermopower, $T$ is temperature, and $\kappa$ is the thermal conductivity, normally the sum of electronic and lattice contributions, $\kappa = \kappa_e + \kappa_l$. The transport functions determining $zT$, and therefore $zT$ itself, are temperature dependent. Recent advances have shown that zT can be enhanced in nanoscale systems by taking advantage of phonon scattering at interfaces to reduce thermal conductivity and quantum confinement and carrier scattering effects to enhance the power factor, $S^2\sigma$ [2]. Currently, reasonably high zT is only found in p-type oxide TE materials. The progress in developing n-type oxide TE materials with comparable zT is still lacking, which calls for more research effort in this area [3]. ZnO-based thermoelectric materials can be one of the examples of this strategy.

## Experimental

Nanostructured sample of ZnO was synthesized via a ball milling for 10 hours using High energy planetary Ballmill. Zinc oxide (>99% purity, Merck, Germany) was used as starting materials. Dry milling of ZnO was carried out in a ball mill equipped with a tungsten carbide balls in at room temperature under atmosphere with a rotation speed of 300 rpm. The ball to powder ratio of 20:1 was employed. The powders were pressed into pellets (cylinder) with diameter of 10mm and thickness of 10 mm using a uni-axial press at a force 2 KN. The specimens were sintered at 200°C for 5 h, and then cooled in the furnace to room temperature. The crystalline phase of the as prepared sample was identified by the powder X-ray diffraction method using Cu Kα radiation (wavelength λ=1.5418 A˚). The thermoelectric parameters were recorded by using the standard techniques. The microstructure of the as-sintered samples was investigated using a Field Emission Scanning Electron Microscope (FE-SEM). Electrical transport of the sample was measured using Physical Property Measurement System (PPMS), Quantum Design Inc., U.S.A. Thermal transport measurement were performed using the thermal transport (TTO) option of PPMS.

## RESULTS AND DISCUSSIONS
### Structural Analysis

The XRD pattern of nanostructured sample of ZnO is shown in Fig.1. The sample is essentially in single phase form. We have not observed any trace of impurity peaks which confirms that it was not affected by the tungsten carbide ball used during ball-milling. All the diffraction peaks can be well indexed to the hexagonal phase ZnO with hexagonal symmetry (space group $P6_3mc$). Using Scherrer's formula (given in eq. 1) one can estimate the particle size which turns out to be ~20nm using the Gaussian fits to the observed maximum intensity X-ray diffraction peak [5].

$$D = \frac{k\lambda}{\beta Cos\theta} - - - - - - - - - - (1)$$

Typical FE-SEM image of the ZnO pellet sintered at 200°C is shown in Fig. 2. It shows that the particles have an almost homogeneous size distribution. This also confirms that the particle size has not increased after sintering the sample at 200°C.

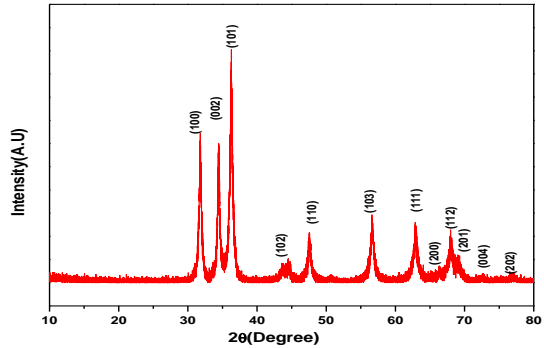

**FIGURE 1.** XRD patterns of the sample ZnO (Ball milled for 10 hours)

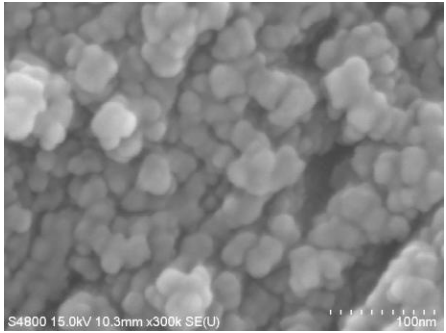

**FIGURE 2**. FE-SEM image of the ZnO pellet sintered at 200°C.

### Thermoelectric Studies

Figure 3 shows the temperature dependence of the Seebeck coefficient, thermal conductivity and electrical resistivity for the sample. The sign of the Seebeck coefficient is negative at the high temperature end indicating that the majority carriers are electrons. The absolute value of Seebeck coefficient of the sample increases with the increase in temperature. This can be explained by using the broadband model [1]. The thermal conductivity decreases significantly in nanostructured sample with respect to bulk zinc oxide. At room temperature $\kappa$ for the nanostructured sample was found to be ~0.75 WK$^{-1}$m$^{-1}$ against typical values of ~37 WK$^{-1}$m$^{-1}$ (by a factor of 50) for bulk samples as reported in the existing literature [6]. Enhancement of phonon scattering could account for the observed reduction of $\kappa$ upon nanostructuring.

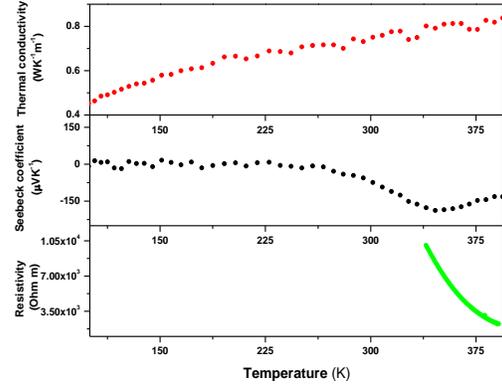

**FIGURE 3.** Temperature dependence of the Seebeck coefficient, thermal conductivity and electrical resistivity for the sample.

## CONCLUSION

We made nanostructured ZnO samples by sintering the ball-milled ZnO. It has a significantly large Seebeck coefficient at high temperatures. Also a large suppression in $\kappa$ is observed compared to the bulk samples. However, we observe that the resistivity of the nanostructured ZnO sample is extremely high. In order to resolve this issue, we plan to make Ag-ZnO nanocomposites where we expect an improved electrical conductivity and hence an enhanced zT. This would be taken up in future studies.

## ACKNOWLEDGMENTS

Authors gratefully acknowledge Department of Atomic Energy, Government of India (Sanction No. 2011/20/37P/03/BRNS/076) for the financial support.

## REFERENCES

1. M. Ohtaki, T. Tsubota and K. Eguchi, J. Appl. Phys. 79, 1816 (1996).
2. S.V. Faleev and F. Léonard, Physical Review B 77, 214304 (2008).
3. J.D. Sugar and D.L. Medlin, Journal of Alloys and Compounds 478, 75 (2009).
4. B.A. Cook, J.L. Harring and C.B. Vining, Journal of Applied Physics 83, 5858 (1998).
5. B. D. Cullity, Elements of X-ray diffraction, 2nd ed., Addison-Wesley series (1978).
6. T. Olorunyolemi, A. Birnboim, Y. Carmel, O. C. Wilson Jr., and I. K. Lloyd, J. Am. Ceram. Soc. 85, 1249 (2002).